\documentclass[preprint,12pt]{elsarticle}
\usepackage{graphics,lscape}

\usepackage{amssymb}

\journal{Nuclear Physics A}

\begin{document}

\begin{frontmatter}


\title{Cross section and reaction rate of $^{92}$Mo(p,$\gamma$)$^{93}$Tc determined from thick target yield measurements}

\author{Gy.~Gy\"urky\corref{cor1}}
\ead{gyurky@atomki.mta.hu}
\cortext[cor1]{corresponding author}
\author{M.~Vakulenko, Zs.~F\"ul\"op, Z.~Hal\'asz, G.G.~Kiss, E.~Somorjai, T.~Sz\"ucs}
\address{Institute for Nuclear Research (MTA Atomki), H-4001 Debrecen, POB.51., Hungary}

\begin{abstract}

For the better understanding of the astrophysical $\gamma$-process the experimental determination of low energy proton- and $\alpha$-capture
cross sections on heavy isotopes is required. The existing data for the $^{92}$Mo(p,$\gamma$)$^{93}$Tc reaction are contradictory and strong fluctuation of the cross section is observed which cannot be explained by the statistical model. In this paper a new determination of the $^{92}$Mo(p,$\gamma$)$^{93}$Tc and $^{98}$Mo(p,$\gamma$)$^{99\rm{m}}$Tc cross sections based on thick target yield measurements are presented and the results are compared with existing data and model calculations. Reaction rates of $^{92}$Mo(p,$\gamma$)$^{93}$Tc at temperatures relevant for the $\gamma$-process are derived directly from the measured thick target yields. The obtained rates are a factor of 2 lower than the ones used in astrophysical network calculations. It is argued that in the case of fluctuating cross sections the thick target yield measurement can be more suited for a reliable reaction rate determination. 
\end{abstract}


\end{frontmatter}


\section{Introduction}
\label{sec:introduction}

The bulk of the chemical elements heavier than iron are thought to be produced by neutron induced reactions in the astrophysical s- and r-processes \cite{kap11,arn07}. A tiny fraction of the heavy elements, however, cannot be synthesized by these processes since some of the most proton rich isotopes (the so-called p nuclei) lie outside the path of both the s- and r-processes. 

It is generally accepted that the main mechanism contributing to the synthesis of the p isotopes is the so-called $\gamma$-process \cite{rau13} which takes place in high temperature stellar environments. The $\gamma$-process converts the isotopes at the bottom of the valley of stability to proton rich species by consecutive ($\gamma$,n) photodisintegration reactions. Charged particle emitting ($\gamma,\alpha$) and ($\gamma$,p) reactions play also an important role in a $\gamma$-process reaction flow model.

The presently available $\gamma$-process models fail to reproduce the p isotope abundances observed in the solar system. This can in part be explained by the ambiguous astrophysical conditions under which the process may take place. On the other hand deficiencies of the nuclear physics input of the models can also be responsible for the failure. 

In lack of experimental data \cite{KADONIS} the reaction rates are typically taken from Hauser-Feshbach statistical model calculations. At the astrophysically relevant low energy range the few available experimental datasets show that the statistical model does not give a satisfactory estimation of the cross sections \cite{rau13}. Especially severe deficiencies are encountered in the case of reactions involving charged particles. Therefore, the measurement of reaction cross sections relevant for the $\gamma$-process is important in order to test the statistical model predictions and to put the $\gamma$-process models on a more reliable ground.

\subsection{The case of the $^{92}$Mo(p,$\gamma$)$^{93}$Tc reaction}
\label{sec:92Mo}

The ($\gamma,\alpha$) and ($\gamma$,p) reactions are preferably studied by measuring the cross section of the inverse capture reactions \cite{rau11}. One of the first reactions that were studied for the purpose of the $\gamma$-process have been the proton capture reactions on several stable Mo isotopes by Sauter and K\"appeler \cite{sau97}. This mass region deserves special attention owing to the exceptionally large natural abundances of $^{92}$Mo and $^{94}$Mo p isotopes which are notoriously and severely underproduced in $\gamma$-process models \cite[and references therein]{rau13,arn03}.

The results of the work of Sauter and K\"appeler \cite{sau97} on the $^{92}$Mo(p,$\gamma$)$^{93}$Tc reaction show a strong fluctuation of the cross section as a function of the energy (see Table 4. and Fig. 5 in \cite{sau97}). This feature is explained by the low level density of the neutron magic $^{93}$Tc isotope. However, one of the basic assumptions of the statistical model is that level density is high enough so that a statistical treatment is possible. The statistical model is thus not able to reproduce the observed strong fluctuations. Therefore, the careful study of this reaction is very important in order to provide reliable reaction rates for $\gamma$-process networks.

Recently the $^{92}$Mo(p,$\gamma$)$^{93}$Tc reaction was studied again by the Cologne group with both activation and in-beam $\gamma$-spectroscopy methods \cite{has10,sau11,sau11b}. No final results of these measurements have been published yet, but the preliminary results show rather big deviation from the data of Ref.\,\cite{sau97} (see Fig. 2. in \cite{sau11b}). In order to investigate the apparent disagreement of the two available datasets and to study the fluctuations of the excitation function, the aim of the present work is to measure  the $^{92}$Mo(p,$\gamma$)$^{93}$Tc cross section using a different experimental technique.

One possible source of uncertainty in cross section measurements utilizing thin targets is the determination of the number of target atoms, the uniformity of the target layers and the stability of the targets under beam bombardment. This uncertainty can be avoided if an infinitely thick target which stops completely the proton beam is used instead of a thin target layer with known thickness. In such a measurement the thick target yield is determined instead of the cross section itself. If the thick target yield as a function of energy is measured with small enough energy steps, the cross section can be obtained by differentiating the yield curve (see Section\,\ref{sec:results} for further details of the cross section determination). The only information which is needed about the target is the stopping power value which is known to a precision of 3.2\,\% in the case of protons in metallic Mo \cite{SRIM-Mo}. 

Radiative proton capture on $^{92}$Mo can populate either the ground state or the isomeric state of $^{93}$Tc. Both of these states decay by electron capture or positron emission to $^{93}$Mo (the isomeric state also decays to the ground state of $^{93}$Tc by internal transition). These decays are followed by characteristic gamma-ray emissions which can be used to determine the partial cross sections leading to the ground and isomeric states separately by the activation method. Table \ref{tab:decay} summarizes the relevant decay parameters of the studied isotopes.

\begin{table*}
\centering
\caption{\label{tab:decay} Decay parameters of the reaction products of the studied reactions. Only those gamma-transitions are listed which are used for the analysis. The data are taken from \cite{NDS92} and \cite{NDS98}.}
\begin{tabular}{lcccc}
\hline
Reaction  & Produced & Half-life  & Gamma-energy & Gamma-intensity\\
& isotope & [hour] & [keV] & [\%] \\
\hline
$^{92}$Mo(p,$\gamma$) & $^{93\rm{g}}$Tc & 2.75\,$\pm$\,0.05 & 1363 & 66.2\,$\pm$\,0.6 \\
                      &            &                   & 1477 & 8.67\,$\pm$\,0.47 \\
                      &            &                   & 1520 & 24.4\,$\pm$\,0.8 \\                        
$^{92}$Mo(p,$\gamma$) & $^{93\rm{m}}$Tc & 0.725\,$\pm$\,0.017 & 392 & 58.9\,$\pm$\,0.9 \\
                      &            &                   & 2645 & 13.3\,$\pm$\,0.6 \\
$^{98}$Mo(p,$\gamma$) & $^{99\rm{m}}$Tc & 6.0067\,$\pm$\,0.0005 & 141 & 89.0\,$\pm$\,0.3 \\
\hline
\end{tabular}
\end{table*}

Using natural isotopic composition targets the cross section of proton induced reactions on heavier Mo isotopes can in principle also be measured with activation. With our experimental conditions only the yield of the $^{98}$Mo(p,$\gamma$)$^{99\rm{m}}$Tc reaction was high enough to be measurable with sufficient precision. This reaction channel was also studied by \cite{sau97}. Owing to the much higher level density of the $^{99}$Tc compound nucleus at the relevant excitation energy range compared to that of $^{93}$Tc\footnote{The large difference in the level densities comes from the fact that $^{93}$Tc is neutron-magic, while $^{99}$Tc is non-magic. Additionally, the excitation energy of the $^{99}$Tc nucleus is about 2.4\,MeV higher at a given proton energy owing to the Q-value difference.}, the cross section of $^{98}$Mo(p,$\gamma$)$^{99\rm{m}}$Tc is a smooth function of energy as observed in \cite{sau97}. The study of this reaction in the present work provides a reliability check of the method.

\section{Experimental procedure}
\label{sec:experimental}

The proton irradiations of 0.5\,mm thick natural isotopic composition metallic Mo targets were done at the 5\,MV Van de Graaff accelerator of the Institute for Nuclear Research (MTA Atomki) in Debrecen, Hungary. The studied energy range between 1700 and 3100\,keV was covered with 50\,keV steps. For each irradiation a fresh Mo target was used. The astrophysically relevant energy range at typical $\gamma$-process temperatures between 2 and 3.5\,GK lies between about 1.4 and 4.0\,MeV \cite{rau10}. The present experiment therefore covers a large part of the relevant energy range. 

The typical beam intensity varied between 1 and 4\,$\mu$A. The beam was wobbled across a surface of 4\,mm in diameter on the target and the target was directly water cooled. The beam intensity was kept as stable as possible, but in order to follow the changes the integrated beam current was recorded in multichannel scaling mode with one minute time basis. The small fluctuations of the beam intensity were taken into account in the analysis. 

The length of the irradiations varied between 20 minutes and 3 hours, the longer irradiations being used at the lower energy region where the cross section is lower. After the irradiations the targets were removed from the chamber and transported to the $\gamma$-counting facility. The typical time between the end of the irradiation and the beginning of the counting was 5 minutes.

\begin{figure}
\centering
\includegraphics[width=\columnwidth]{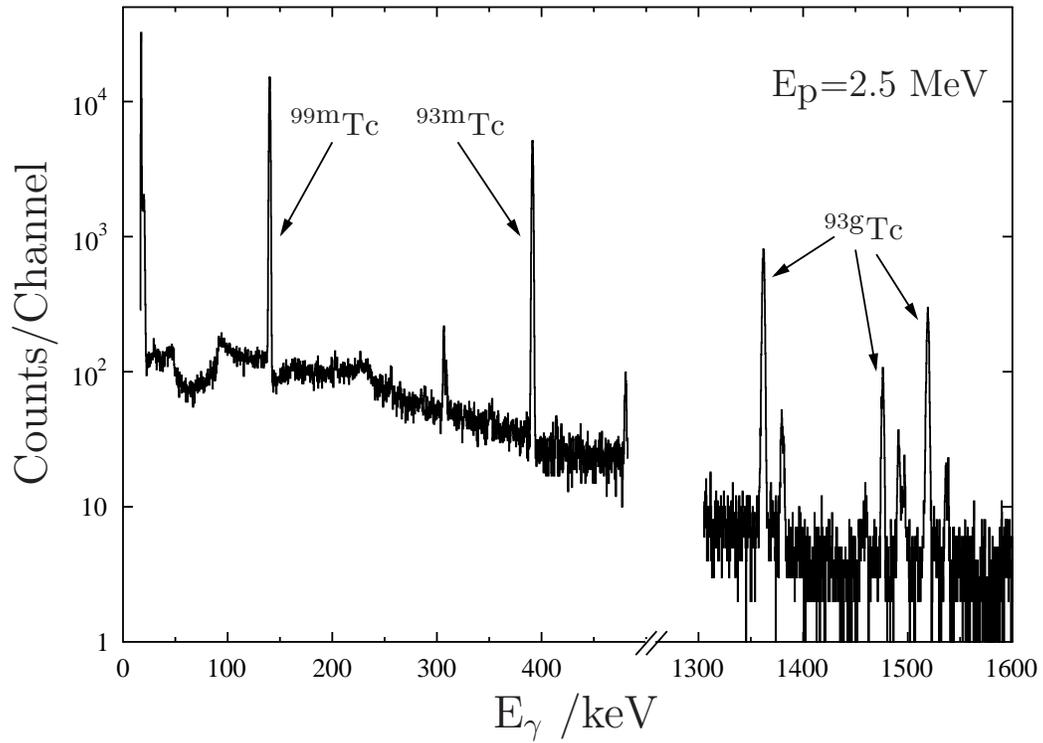}
\caption{\label{fig:spectrum} Typical $\gamma$-spectrum measured on a target irradiated with a proton beam of 2.5\,MeV. The channel width is 0.28\,keV. The different $\gamma$-peaks used for the analysis are indicated. The peaks not labeled are from laboratory background, from the decay of not studied Tc isotopes (like $^{101}$Tc) or from true coincidence summing effect between X-rays and $\gamma$-rays.}
\end{figure}

\begin{figure}
\centering
\includegraphics[width=0.8\columnwidth]{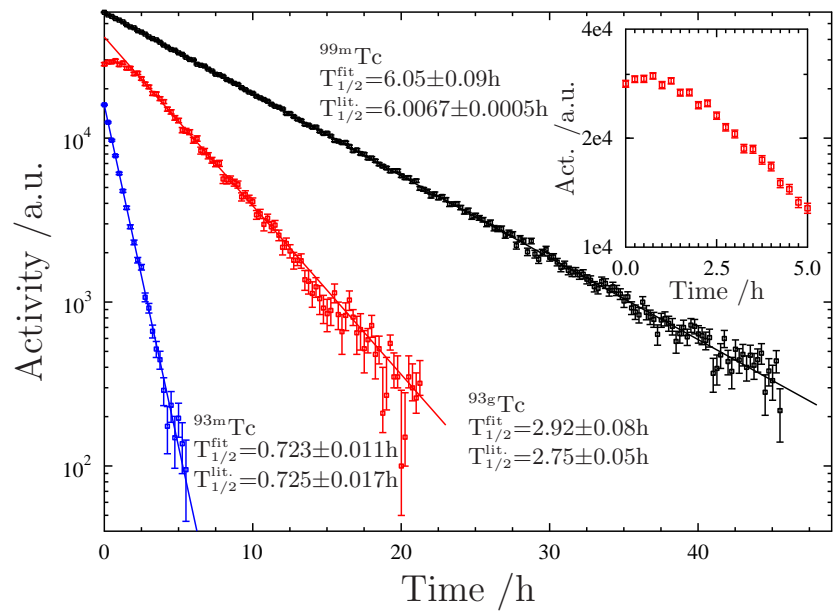}
\caption{\label{fig:decay} Decay curves of the three produced isotopes measured on a target irradiated with a 2.8\,MeV proton beam. In the case of the $^{93\rm{g}}$Tc isotopes the deviation form the exponential function near the beginning of the counting (shown enlarged in the inset) is due to the feeding of the ground state by the shorter lived isomer. The deduced half-lives from an exponential fit as well as the literature half-life values are also shown.}
\end{figure}

The $\gamma$-radiation following the decay of the reaction products was measured with a 100\,\% relative efficiency HPGe detector placed in a low background complete 4$\pi$ shielding. The absolute efficiency of the detector in the counting geometry was measured with several calibrated radioactive sources following a procedure detailed in Ref\,\cite{hal12}. The length of the countings varied between half an hour and 15 hours and the spectra were stored regularly in order to follow the decay of the different reaction products. Figure\,\ref{fig:spectrum} shows a typical $\gamma$-spectrum indicating the $\gamma$-lines used for the analysis while in fig.\,\ref{fig:decay} typical decay curves of the three produced isotopes are shown. The obtained half-lives from a fit to the data are in good agreement with the literature data which proves that there is no contamination in the studied peaks.

\section{Results and discussion}
\label{sec:results}

\begin{table}
\centering
\caption{\label{tab:92yield} Thick target yield of the $^{92}$Mo(p,$\gamma$)$^{93}$Tc reaction. The quoted uncertainties are statistical only.}
\begin{tabular}{crclrclrcl}
\hline
E$_p$ & \multicolumn{3}{c}{Ground state} & \multicolumn{3}{c}{Isomeric state} & \multicolumn{3}{c}{Total}\\
\cline{2-10}
[keV] & \multicolumn{9}{c}{10$^{-12}$ reactions/projectile}\\
\hline
1700	&	0.401	&	$\pm$	&	0.029	&	0.575	&	$\pm$	&	0.025	&	0.976	&	$\pm$	&	0.038	\\
1750	&	0.454	&	$\pm$	&	0.024	&	0.623	&	$\pm$	&	0.020	&	1.08	&	$\pm$	&	0.03	\\
1800	&	1.006	&	$\pm$	&	0.042	&	1.48	&	$\pm$	&	0.03	&	2.48	&	$\pm$	&	0.05	\\
1850	&	1.26	&	$\pm$	&	0.05	&	2.13	&	$\pm$	&	0.04	&	3.38	&	$\pm$	&	0.06	\\
1900	&	1.73	&	$\pm$	&	0.06	&	2.96	&	$\pm$	&	0.05	&	4.69	&	$\pm$	&	0.08	\\
1950	&	2.82	&	$\pm$	&	0.09	&	4.39	&	$\pm$	&	0.06	&	7.21	&	$\pm$	&	0.11	\\
2000	&	3.67	&	$\pm$	&	0.12	&	5.76	&	$\pm$	&	0.08	&	9.43	&	$\pm$	&	0.14	\\
2050	&	4.97	&	$\pm$	&	0.30	&	9.04	&	$\pm$	&	0.19	&	14.0	&	$\pm$	&	0.4	\\
2100	&	6.43	&	$\pm$	&	0.29	&	10.3	&	$\pm$	&	0.2	&	16.7	&	$\pm$	&	0.3	\\
2150	&	7.58	&	$\pm$	&	0.27	&	12.8	&	$\pm$	&	0.2	&	20.4	&	$\pm$	&	0.3	\\
2200	&	13.2	&	$\pm$	&	0.4	&	19.2	&	$\pm$	&	0.2	&	32.3	&	$\pm$	&	0.5	\\
2250	&	16.1	&	$\pm$	&	0.5	&	22.8	&	$\pm$	&	0.3	&	38.8	&	$\pm$	&	0.6	\\
2300	&	18.9	&	$\pm$	&	0.6	&	27.1	&	$\pm$	&	0.3	&	46.0	&	$\pm$	&	0.7	\\
2350	&	21.6	&	$\pm$	&	0.7	&	35.5	&	$\pm$	&	0.3	&	57.2	&	$\pm$	&	0.7	\\
2400	&	25.4	&	$\pm$	&	0.5	&	41.6	&	$\pm$	&	0.4	&	67.1	&	$\pm$	&	0.6	\\
2450	&	32.1	&	$\pm$	&	1.0	&	55.7	&	$\pm$	&	0.5	&	87.8	&	$\pm$	&	1.1	\\
2500	&	42.9	&	$\pm$	&	1.1	&	70.6	&	$\pm$	&	0.5	&	114	&	$\pm$	&	1.3	\\
2550	&	49.4	&	$\pm$	&	1.4	&	87.8	&	$\pm$	&	0.7	&	137	&	$\pm$	&	1.5	\\
2600	&	60.7	&	$\pm$	&	1.5	&	102	&	$\pm$	&	0.7	&	163	&	$\pm$	&	1.7	\\
2650	&	72.4	&	$\pm$	&	1.7	&	121	&	$\pm$	&	0.8	&	193	&	$\pm$	&	1.9	\\
2700	&	81.2	&	$\pm$	&	1.8	&	139	&	$\pm$	&	0.9	&	220	&	$\pm$	&	2.0	\\
2750	&	87.2	&	$\pm$	&	1.0	&	150	&	$\pm$	&	0.5	&	238	&	$\pm$	&	1.2	\\
2800	&	106	&	$\pm$	&	1.5	&	175	&	$\pm$	&	0.6	&	281	&	$\pm$	&	1.7	\\
2850	&	115	&	$\pm$	&	1.4	&	204	&	$\pm$	&	0.7	&	319	&	$\pm$	&	1.5	\\
2900	&	133	&	$\pm$	&	1.6	&	231	&	$\pm$	&	0.8	&	364	&	$\pm$	&	1.8	\\
2950	&	151	&	$\pm$	&	1.7	&	249	&	$\pm$	&	0.8	&	401	&	$\pm$	&	1.8	\\
3000	&	176	&	$\pm$	&	1.8	&	281	&	$\pm$	&	0.9	&	456	&	$\pm$	&	2.0	\\
3050	&	208	&	$\pm$	&	2.0	&	326	&	$\pm$	&	1.0	&	533	&	$\pm$	&	2.2	\\
3100	&	243	&	$\pm$	&	1.8	&	377	&	$\pm$	&	1.0	&	620	&	$\pm$	&	2.1	\\
\hline
\end{tabular}
\end{table}

\begin{table}
\centering
\caption{\label{tab:98yield} Thick target yield of the $^{98}$Mo(p,$\gamma$)$^{99\rm{m}}$Tc reaction. The quoted uncertainties are statistical only. The yield values are given in 10$^{-12}$ reactions/projectile.}
\begin{tabular}{crclcrcl}
\hline
E$_p$ & \multicolumn{3}{c}{$^{98}$Mo(p,$\gamma$)$^{99\rm{m}}$Tc} & E$_p$ & \multicolumn{3}{c}{$^{98}$Mo(p,$\gamma$)$^{99\rm{m}}$Tc}  \\
\,[keV] & \multicolumn{3}{c}{yield} & \,[keV] & \multicolumn{3}{c}{yield}\\
\hline
1700	&	0.762	&	$\pm$	&	0.011	&	2450	&	164	&	$\pm$	&	0.9	\\
1750	&	1.17	&	$\pm$	&	0.01	&	2500	&	212	&	$\pm$	&	1.0	\\
1800	&	1.90	&	$\pm$	&	0.02	&	2550	&	276	&	$\pm$	&	1.3	\\
1850	&	2.89	&	$\pm$	&	0.02	&	2600	&	353	&	$\pm$	&	1.5	\\
1900	&	4.25	&	$\pm$	&	0.02	&	2650	&	398	&	$\pm$	&	1.6	\\
1950	&	6.28	&	$\pm$	&	0.04	&	2700	&	435	&	$\pm$	&	1.7	\\
2000	&	9.20	&	$\pm$	&	0.07	&	2750	&	454	&	$\pm$	&	1.1	\\
2050	&	14.0	&	$\pm$	&	0.2	&	2800	&	469	&	$\pm$	&	0.8	\\
2100	&	19.5	&	$\pm$	&	0.2	&	2850	&	483	&	$\pm$	&	1.4	\\
2150	&	27.3	&	$\pm$	&	0.1	&	2900	&	496	&	$\pm$	&	1.6	\\
2200	&	38.0	&	$\pm$	&	0.3	&	2950	&	507	&	$\pm$	&	1.5	\\
2250	&	52.3	&	$\pm$	&	0.3	&	3000	&	517	&	$\pm$	&	1.5	\\
2300	&	70.1	&	$\pm$	&	0.4	&	3050	&	531	&	$\pm$	&	1.6	\\
2350	&	95.1	&	$\pm$	&	0.5	&	3100	&	546	&	$\pm$	&	1.1	\\
2400	&	126	&	$\pm$	&	0.3	&								\\
\hline
\end{tabular}
\end{table}

The obtained thick target yields for the studied reactions are listed in Tables \ref{tab:92yield} and \ref{tab:98yield}. The yield is defined here as the number of reactions per projectile supposing a chemically pure natural isotopic composition Mo target with infinite thickness. The uncertainty of the proton beam energy is about 2\,keV. The quoted uncertainties of the yield values are statistical only. In order to obtain the total uncertainties, the following systematic errors must be added quadratically to the relative statistical uncertainties: current integration (3\,\%), detector efficiency (4\,\%), relative intensities of decay $\gamma$-rays (1-3\,\%, see table \ref{tab:decay}). These uncertainties are common for all the individual measurements, so they can be treated separately when making the differentiation for cross section determination (see below). The contribution of the half-life error to the final uncertainty depends on the actual length of the counting. These errors are, however, typically below or around 2\,\%, and therefore do not give a significant contribution to the final uncertainty. 

Before discussing the cross section derived from the thick target yield, it should be mentioned that the thick target yield of the reactions studied in the present work was already measured by N.A. Roughton {\it et al.} in 1979 \cite{rou79}. In this work many reactions were studied and there is limited information about the experiment and data analysis. There is a relatively good agreement between these data and the present work in the case of the $^{92}$Mo(p,$\gamma$)$^{93}$Tc reaction. For the $^{98}$Mo(p,$\gamma$)$^{99\rm{m}}$Tc reaction, however, the yield obtained in \cite{rou79} is about two orders of magnitude higher than in the present work. Such a high yield seems very unlikely since the results of the present work agrees relatively well with that of \cite{sau97} and with all theoretical calculations (see below). Since no cross section but astrophysical reaction rate was derived from the thick target yield, the results of N.A. Roughton {\it et al.} \cite{rou79} are only included in the discussion of reaction rates in Sec.\,\ref{sec:rate}.

The thick target yield $Y(E)$ as a function of proton energy $E$ is related to the reaction cross section $\sigma(E)$ by the following formula:

\begin{equation}
\label{eq:yield}
	Y(E)=\int_0^E \frac{\sigma(E')}{\varepsilon(E')}dE'
\end{equation}
where $\varepsilon(E)$ is the effective stopping power for the studied isotope, i.e. the stopping power of chemically pure Mo multiplied by the isotopic abundance of the studied isotope. If the thick target yield is measured at energies $E$ and $E+\Delta E$, the average cross section between $E$ and $E+\Delta E$ can be obtained by subtracting the two yields:

\begin{equation}
	\sigma(E;E+\Delta E)=\frac{[Y(E+\Delta E)-Y(E)] \cdot \varepsilon(E;E+\Delta E)}{\Delta E}
\end{equation}
Decreasing the step size would allow a more detailed mapping of the excitation function.
A too small value of $\Delta E$, however,  would result in the subtraction of two similar numbers, leading to an increased statistical uncertainty. As a compromise, in the present work the thick target yield was measured in 50\,keV steps, so $\Delta E$\,=\,50\,keV. 

\subsection{The $^{98}$Mo(p,$\gamma$)$^{99\rm{m}}$Tc reaction}

Let us start the discussion with the $^{98}$Mo(p,$\gamma$)$^{99\rm{m}}$Tc reaction. In Table~\ref{tab:98sigma} the cross section values obtained from the thick target yield are listed. The energy corresponding to each extracted cross section value is taken at the center of the relevant energy bin. The $\pm$\,25\,keV uncertainty of the energy reflects the 50\,keV energy step size of the thick target yield curve. The quoted uncertainties of the cross section are statistical only. In order to get the total uncertainty, a systematic error of 7\,\% must be added in quadrature to the values. In addition to the components listed above, this 7\,\% uncertainty comprises now also the uncertainty of the stopping power. The stopping power uncertainty does not affect the measured thick target yield, but through the above formulas influences the cross section. According to the SRIM tables \cite{SRIM-Mo}, a 3.2\,\% uncertainty is assigned to the stopping of protons in molybdenum.

The relative uncertainties of the highest energy points are higher than at low energies. The reason is that the cross section drops above 2.65\,MeV (where the $^{98}$Mo(p,n)$^{98}$Tc reaction channel opens) which results in a thick target yield function increasing slowly. The subtraction of two similar numbers causes thus the increased statistical uncertainty.  

\begin{figure}
\centering
\includegraphics[width=\columnwidth]{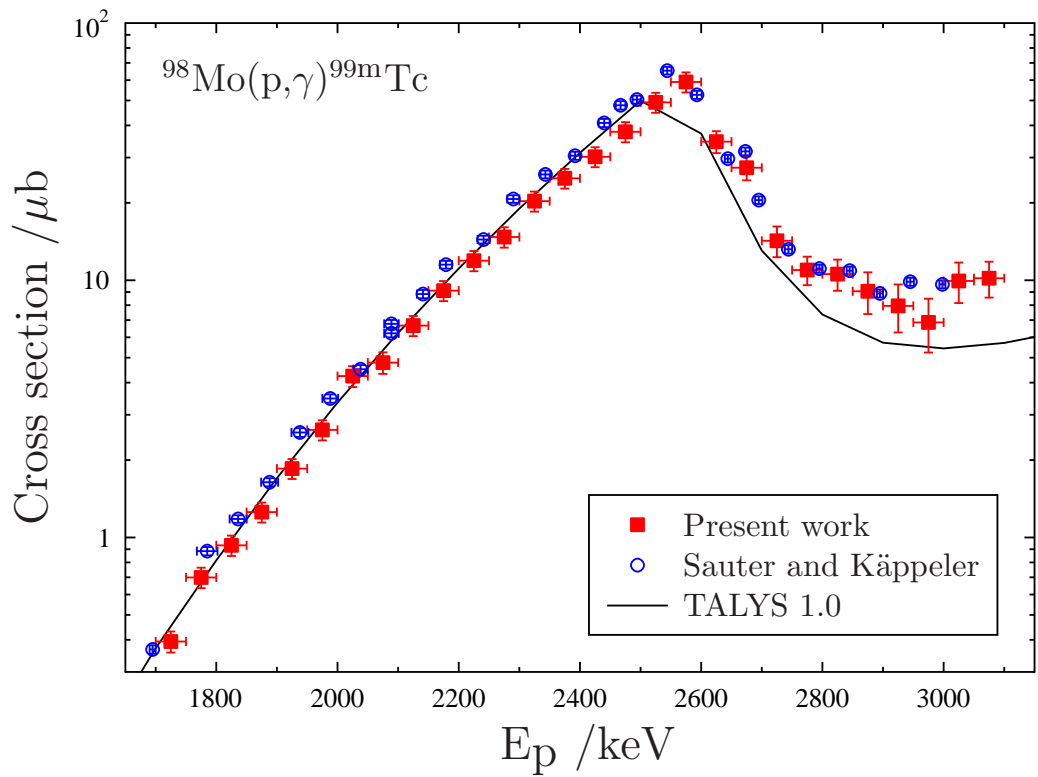}
\caption{\label{fig:98Mo_res} Cross section of the $^{98}$Mo(p,$\gamma$)$^{99\rm{m}}$Tc reaction. The results of the present work obtained from thick target yield measurement is compared with the data of Sauter and K\"appeler \cite{sau97} and the prediction of the statistical model code TALYS 1.0 \cite{TALYS}.}
\end{figure}

The obtained cross sections are compared with the data of Sauter and K\"appeler \cite{sau97} in Fig.\,\ref{fig:98Mo_res}. The result of statistical model calculations using the TALYS code \cite{TALYS} is also included in the figure\footnote{Version 1.0 of the TALYS code was used with its default input parameters. The most relevant input parameters are: optical potential of Koning and Delaroche \cite{kon03}, the Brink-Axel $\gamma$-ray strength function \cite{bri57} and the phenomenological constant temperature + Fermi gas model of Gilbert and Cameron \cite{gil65} for the level density.}. Excellent agreement is found between the two experimental datasets which proves the reliability of the method of obtaining precise cross sections from thick target yield measurements. 

The cross section is a smoothly varying function of the energy and the statistical model calculation gives a good description of the measured data (with some small deviation at high energies). This indicates that the level density in $^{99}$Tc is high enough so that no fluctuations in the cross section is observed and the applicability criteria of the statistical model are fulfilled.

\begin{table}
\centering
\caption{\label{tab:98sigma} Cross section of the $^{98}$Mo(p,$\gamma$)$^{99\rm{m}}$Tc reaction. The quoted uncertainties are statistical only.}
\begin{tabular}{rclrclrclrcl}
\hline
\multicolumn{3}{c}{E$_p$} & \multicolumn{3}{c}{cross section} & \multicolumn{3}{c}{E$_p$} & \multicolumn{3}{c}{cross section}\\
\multicolumn{3}{c}{[keV]} & \multicolumn{3}{c}{[$\mu$barn]} & \multicolumn{3}{c}{[keV]} & \multicolumn{3}{c}{[$\mu$barn]}\\
\hline
1725	&	$\pm$	&	25	&	0.394	&	$\pm$	&	0.015	&	2425	&	$\pm$	&	25	&	30.2	&	$\pm$	&	0.7	\\
1775	&	$\pm$	&	25	&	0.699	&	$\pm$	&	0.021	&	2475	&	$\pm$	&	25	&	37.8	&	$\pm$	&	1.1	\\
1825	&	$\pm$	&	25	&	0.932	&	$\pm$	&	0.030	&	2525	&	$\pm$	&	25	&	49.2	&	$\pm$	&	1.3	\\
1875	&	$\pm$	&	25	&	1.25	&	$\pm$	&	0.03	&	2575	&	$\pm$	&	25	&	59.0	&	$\pm$	&	1.5	\\
1925	&	$\pm$	&	25	&	1.85	&	$\pm$	&	0.04	&	2625	&	$\pm$	&	25	&	34.6	&	$\pm$	&	1.6	\\
1975	&	$\pm$	&	25	&	2.62	&	$\pm$	&	0.07	&	2675	&	$\pm$	&	25	&	27.4	&	$\pm$	&	1.8	\\
2025	&	$\pm$	&	25	&	4.24	&	$\pm$	&	0.16	&	2725	&	$\pm$	&	25	&	14.2	&	$\pm$	&	1.5	\\
2075	&	$\pm$	&	25	&	4.78	&	$\pm$	&	0.21	&	2775	&	$\pm$	&	25	&	11.0	&	$\pm$	&	1.0	\\
2125	&	$\pm$	&	25	&	6.67	&	$\pm$	&	0.19	&	2825	&	$\pm$	&	25	&	10.6	&	$\pm$	&	1.2	\\
2175	&	$\pm$	&	25	&	9.11	&	$\pm$	&	0.24	&	2875	&	$\pm$	&	25	&	9.06	&	$\pm$	&	1.48	\\
2225	&	$\pm$	&	25	&	11.9	&	$\pm$	&	0.3	&	2925	&	$\pm$	&	25	&	7.95	&	$\pm$	&	1.54	\\
2275	&	$\pm$	&	25	&	14.7	&	$\pm$	&	0.4	&	2975	&	$\pm$	&	25	&	6.86	&	$\pm$	&	1.51	\\
2325	&	$\pm$	&	25	&	20.3	&	$\pm$	&	0.5	&	3025	&	$\pm$	&	25	&	9.94	&	$\pm$	&	1.57	\\
2375	&	$\pm$	&	25	&	24.9	&	$\pm$	&	0.5	&	3075	&	$\pm$	&	25	&	10.2	&	$\pm$	&	1.4	\\
\hline
\end{tabular}
\end{table}

\subsection{The $^{92}$Mo(p,$\gamma$)$^{93}$Tc reaction}

The cross section results for the $^{92}$Mo(p,$\gamma$)$^{93}$Tc reaction are listed in Tab.\,\ref{tab:92sigma}. The quoted cross section uncertainties are again statistical only, the total uncertainty can be obtained with the quadratic addition of a total 7\,\% systematic uncertainty. Partial cross section leading to the ground and isomeric state of $^{93}$Tc as well as total cross sections are listed separately. 

\begin{figure}
\centering
\includegraphics[width=\columnwidth]{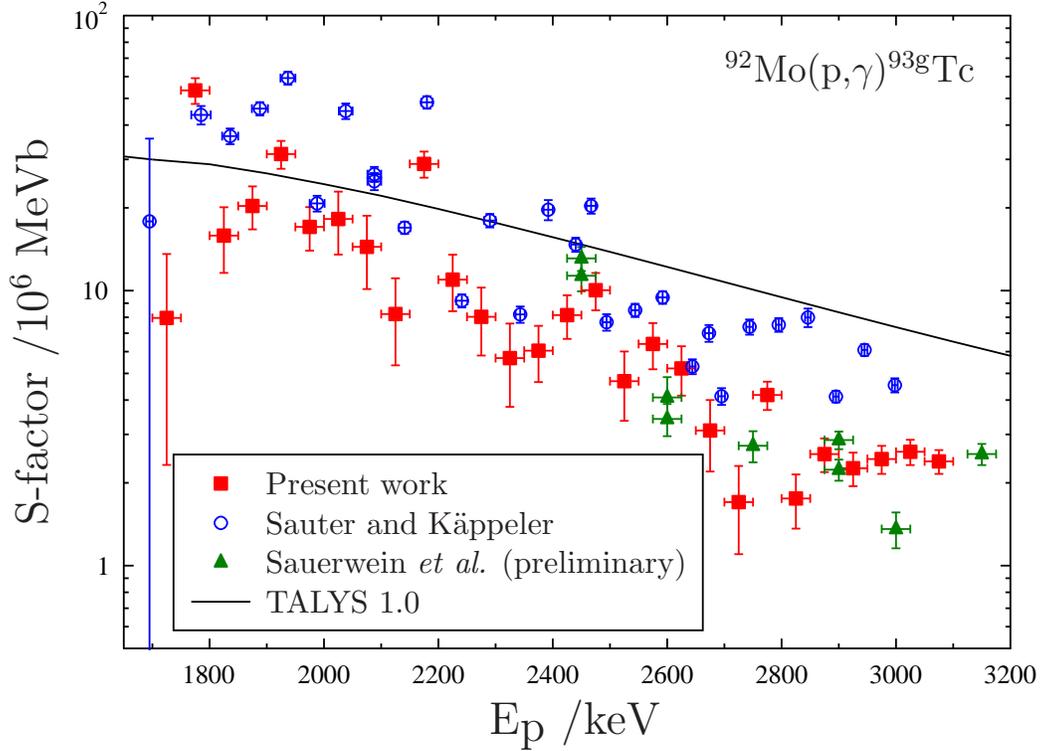}
\caption{\label{fig:92gMo_res} Astrophysical S-factor of the $^{92}$Mo(p,$\gamma$)$^{93\rm{g}}$Tc reaction. The results of the present work obtained from thick target yield measurement is compared with the data of Sauter and K\"appeler \cite{sau97}, with the preliminary results of the Cologne group \cite{has10,sau11,sau11b} and with the prediction of the statistical model code TALYS 1.0 \cite{TALYS}. The original Sauter and K\"appeler \cite{sau97} data are plotted without using the recommended normalization with a factor of 0.85.}
\end{figure}

\begin{figure}
\centering
\includegraphics[width=\columnwidth]{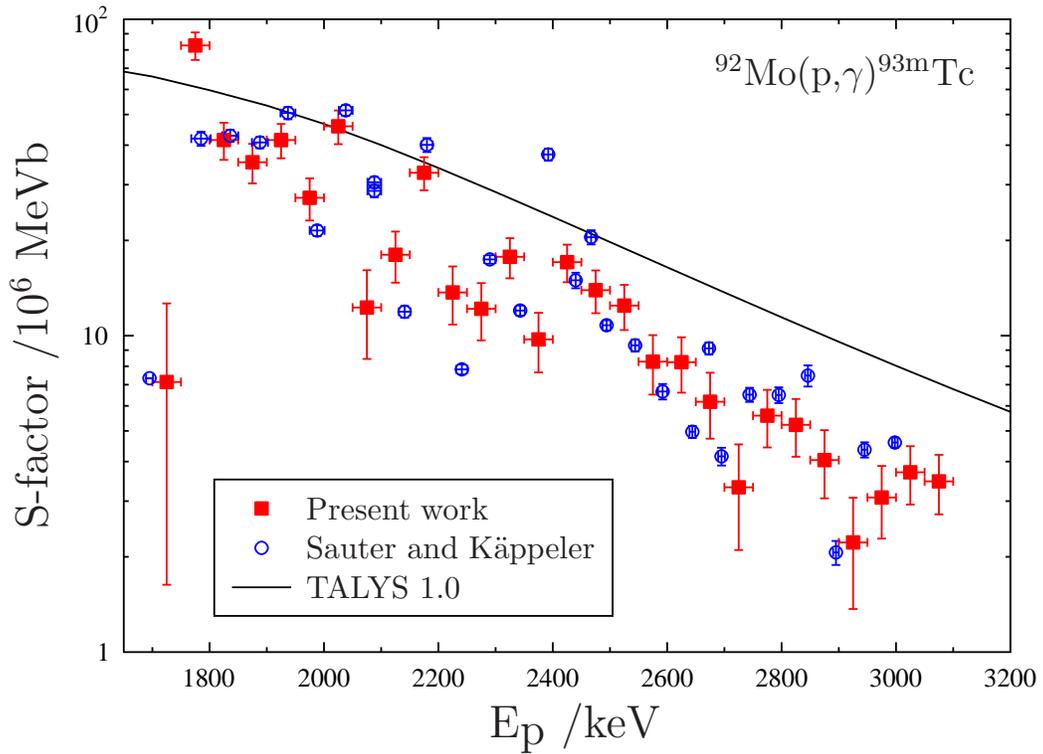}
\caption{\label{fig:92mMo_res} Astrophysical S-factor of the $^{92}$Mo(p,$\gamma$)$^{93\rm{m}}$Tc reaction. The results of the present work obtained from thick target yield measurement is compared with the data of \cite{sau97} and the prediction of the statistical model code TALYS 1.0 \cite{TALYS}.}
\end{figure}

\begin{figure}
\centering
\includegraphics[width=\columnwidth]{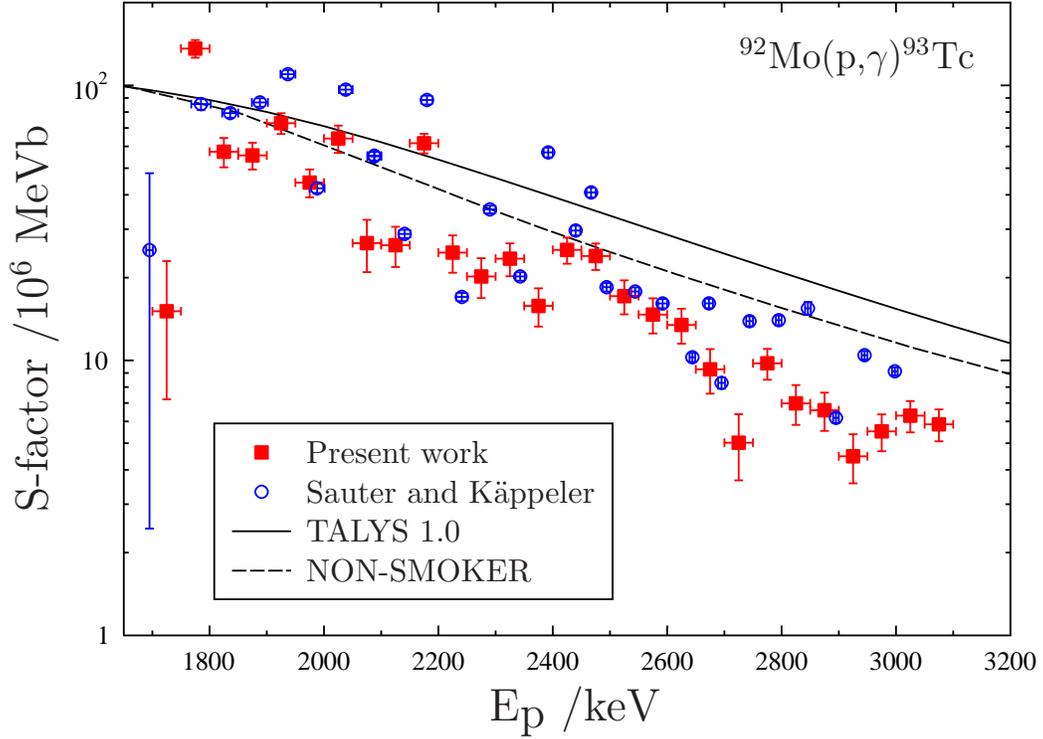}
\caption{\label{fig:92totMo_res} Total astrophysical S-factor of the $^{92}$Mo(p,$\gamma$)$^{93}$Tc reaction. The results of the present work obtained from thick target yield measurement is compared with the data of \cite{sau97} and the prediction of the statistical model codes TALYS 1.0 \cite{TALYS} and NON-SMOKER \cite{NONSMOKER}. The original Sauter and K\"appeler \cite{sau97} data are plotted without using the recommended normalization of the ground state cross section with a factor of 0.85.}
\end{figure}

Figures \ref{fig:92gMo_res}, \ref{fig:92mMo_res} and \ref{fig:92totMo_res} show the  ground state, isomeric state and total cross sections, respectively. Since the cross section changes almost three orders of magnitiude in the studied energy range, the results are plotted in the form of less energy dependent astrophysical S-factor \cite{ili07} in order to observe the details more easily. Sauter and K\"appeler \cite{sau97} also measured separately the cross sections to the ground and isomeric state, thus these results are also included in the figures for comparison. The Cologne group \cite{has10,sau11,sau11b}, on the other hand, published preliminary results only for the ground state cross section, which is shown in Fig.\,\ref{fig:92gMo_res}. The predictions of statistical model code TALYS \cite{TALYS} are also shown in the figures. For the total cross section the calculations obtained with the NON-SMOKER \cite{NONSMOKER} code\footnote{The NON-SMOKER code as published in Ref.\,\cite{NONSMOKER} uses microscopic optical potential of Jeukenne, Lejeune and Mahaux \cite{JLM} and level density based on the back-shifted Fermi-gas formalism by Rauscher, Thielemann and Kratz \cite{rau97}. For the treatment of further input parameters see the original publication \cite{NONSMOKER}.}, which is widely used in astrophysical reaction network calculations, are also shown. This code does not provide data separately for the cross section to the ground and isomeric state (the $\gamma$-cascades are not followed explicitely to the low lying states and ground state), therefore the NON-SMOKER \cite{NONSMOKER} results are shown only in Fig.\,\ref{fig:92totMo_res}. 

The present results confirm the observations of Sauter and K\"appeler \cite{sau97} that the cross section shows strong fluctuations as a function of energy. This indicates that the level density of the neutron magic $^{93}$Tc isotope is indeed low and therefore the conditions of the statistical model are not fulfilled. The inherent feature of the present thick target yield measurement is that it averages out any fluctuation of the cross section over 50\,keV energy intervals. Therefore, it cannot be expected that the results are in agreement with that of Sauter and K\"appeler \cite{sau97} since in that case the cross section averaging energy interval owing to the target thickness was only between about 9 and 40 keV. However, the gross feature of the two datasets (e.g. the positions of some high and low points in the excitation function) are in relatively good agreement.

In the case of the ground state cross section the results of Sauter and K\"appeler \cite{sau97} are systematically higher than the present work, while the preliminary results of the Cologne group \cite{has10,sau11,sau11b} are in good agreement with the present data. It should be noted, however, that Sauter and K\"appeler \cite{sau97} used different decay parameters for their activation measurement. See the difference of Table~\ref{tab:decay} in the present paper and Table~I in Ref.~\cite{sau97} for $^{93\rm{g}}$Tc. An older nuclear data evaluation \cite{NDS92old} was used by \cite{sau97} than in the present work \cite{NDS92} and different relative gamma-ray intensities are listed in the two tables. The two evaluations are based on the same experimental data and therefore the reason of the differences is not fully understood. In order to update to the latest data evaluation and to be able to compare the different $^{92}$Mo(p,$\gamma$)$^{93\rm{g}}$Tc experiment, we recommend to multiply the Sauter and K\"appeler $^{92}$Mo(p,$\gamma$)$^{93\rm{g}}$Tc cross section data by a factor of 0.85. Such a downscaling brings the Sauter and K\"appeler data closer to our results and the results of the Cologne group, but some disagreement still remains.

In the case of the isomeric state cross section there is a relatively good overall agreement between the present data and that of Sauter and K\"appeler. The comparison of the total cross section reflects the difference in the ground state cross section: the Sauter and K\"appeler data are systematically higher than the present results. The difference is, however, not strong owing to the fact that the isomeric cross section is the dominant over the ground state one. On average, the statistical model calculations using both the TALYS \cite{TALYS} and NON-SMOKER \cite{NONSMOKER} codes overestimate the measured total cross sections by a factor of about 2 to 3. 

Knowing the fluctuating nature of the excitation function, the thick target yield measurement method applied in the present work has a significant advantage over the direct thin-target cross section measurement in astrophysical applications. With a thin target cross section measurement most of the studied energy range is actually not probed. Strong maxima or minima in the excitation function can be accidentally missed by not having the right proton energy. In the case of a thick target measurement, on the other hand, the whole energy range is mapped without missing any part of it. Since in a stellar environment nuclei with continuous energy distribution react, more reliable reaction rate can be obtained starting from a thick target yield measurement. This is elaborated further in the next section.

\begin{table}
\centering
\caption{\label{tab:92sigma} Cross section of the $^{92}$Mo(p,$\gamma$)$^{93}$Tc reaction. The quoted uncertainties are statistical only.}
\begin{tabular}{rclrclrclrcl}
\hline
\multicolumn{3}{c}{E$_p$} & \multicolumn{3}{c}{Ground state} & \multicolumn{3}{c}{Isomeric state} & \multicolumn{3}{c}{Total}\\
\cline{4-12}
\multicolumn{3}{c}{[keV]} & \multicolumn{9}{c}{[$\mu$barn]}\\
\hline
1725	&	$\pm$	&	25	&	0.084	&	$\pm$	&	0.060	&	0.076	&	$\pm$	&	0.058	&	0.160	&	$\pm$	&	0.083	\\
1775	&	$\pm$	&	25	&	0.864	&	$\pm$	&	0.075	&	1.34	&	$\pm$	&	0.075	&	2.20	&	$\pm$	&	0.11	\\
1825	&	$\pm$	&	25	&	0.383	&	$\pm$	&	0.100	&	1.00	&	$\pm$	&	0.107	&	1.39	&	$\pm$	&	0.15	\\
1875	&	$\pm$	&	25	&	0.722	&	$\pm$	&	0.120	&	1.26	&	$\pm$	&	0.146	&	1.98	&	$\pm$	&	0.19	\\
1925	&	$\pm$	&	25	&	1.62	&	$\pm$	&	0.16	&	2.14	&	$\pm$	&	0.20	&	3.75	&	$\pm$	&	0.25	\\
1975	&	$\pm$	&	25	&	1.25	&	$\pm$	&	0.21	&	2.00	&	$\pm$	&	0.26	&	3.26	&	$\pm$	&	0.33	\\
2025	&	$\pm$	&	25	&	1.89	&	$\pm$	&	0.47	&	4.75	&	$\pm$	&	0.43	&	6.64	&	$\pm$	&	0.64	\\
2075	&	$\pm$	&	25	&	2.08	&	$\pm$	&	0.60	&	1.77	&	$\pm$	&	0.53	&	3.84	&	$\pm$	&	0.80	\\
2125	&	$\pm$	&	25	&	1.63	&	$\pm$	&	0.56	&	3.56	&	$\pm$	&	0.58	&	5.19	&	$\pm$	&	0.81	\\
2175	&	$\pm$	&	25	&	7.76	&	$\pm$	&	0.69	&	8.79	&	$\pm$	&	0.77	&	16.6	&	$\pm$	&	1.0	\\
2225	&	$\pm$	&	25	&	3.96	&	$\pm$	&	0.89	&	4.95	&	$\pm$	&	0.95	&	8.91	&	$\pm$	&	1.30	\\
2275	&	$\pm$	&	25	&	3.86	&	$\pm$	&	1.04	&	5.84	&	$\pm$	&	1.10	&	9.70	&	$\pm$	&	1.51	\\
2325	&	$\pm$	&	25	&	3.60	&	$\pm$	&	1.19	&	11.2	&	$\pm$	&	1.34	&	14.8	&	$\pm$	&	1.8	\\
2375	&	$\pm$	&	25	&	5.00	&	$\pm$	&	1.11	&	8.05	&	$\pm$	&	1.59	&	13.1	&	$\pm$	&	1.9	\\
2425	&	$\pm$	&	25	&	8.73	&	$\pm$	&	1.48	&	18.3	&	$\pm$	&	2.0	&	27.0	&	$\pm$	&	2.5	\\
2475	&	$\pm$	&	25	&	13.8	&	$\pm$	&	2.0	&	19.2	&	$\pm$	&	2.5	&	33.0	&	$\pm$	&	3.2	\\
2525	&	$\pm$	&	25	&	8.22	&	$\pm$	&	2.26	&	21.8	&	$\pm$	&	3.1	&	30.1	&	$\pm$	&	3.8	\\
2575	&	$\pm$	&	25	&	14.2	&	$\pm$	&	2.6	&	18.4	&	$\pm$	&	3.6	&	32.6	&	$\pm$	&	4.4	\\
2625	&	$\pm$	&	25	&	14.6	&	$\pm$	&	2.8	&	23.0	&	$\pm$	&	4.2	&	37.6	&	$\pm$	&	5.0	\\
2675	&	$\pm$	&	25	&	10.8	&	$\pm$	&	3.1	&	21.5	&	$\pm$	&	4.8	&	32.3	&	$\pm$	&	5.7	\\
2725	&	$\pm$	&	25	&	7.36	&	$\pm$	&	2.56	&	14.3	&	$\pm$	&	5.1	&	21.7	&	$\pm$	&	5.7	\\
2775	&	$\pm$	&	25	&	22.3	&	$\pm$	&	2.2	&	29.8	&	$\pm$	&	5.6	&	52.0	&	$\pm$	&	6.1	\\
2825	&	$\pm$	&	25	&	11.5	&	$\pm$	&	2.5	&	34.2	&	$\pm$	&	6.5	&	45.6	&	$\pm$	&	6.9	\\
2875	&	$\pm$	&	25	&	20.3	&	$\pm$	&	2.5	&	32.2	&	$\pm$	&	7.4	&	52.5	&	$\pm$	&	7.8	\\
2925	&	$\pm$	&	25	&	21.9	&	$\pm$	&	2.7	&	21.5	&	$\pm$	&	8.1	&	43.3	&	$\pm$	&	8.5	\\
2975	&	$\pm$	&	25	&	28.5	&	$\pm$	&	2.8	&	35.9	&	$\pm$	&	8.8	&	64.4	&	$\pm$	&	9.2	\\
3025	&	$\pm$	&	25	&	36.4	&	$\pm$	&	3.1	&	51.9	&	$\pm$	&	10.0	&	88.3	&	$\pm$	&	10.4	\\
3075	&	$\pm$	&	25	&	40.1	&	$\pm$	&	3.1	&	58.1	&	$\pm$	&	11.4	&	98.2	&	$\pm$	&	11.8	\\
\hline
\end{tabular}
\end{table}

\section{Astrophysical reaction rate from thick target yields}
\label{sec:rate}

The thermonuclear reaction rate\footnote{In this paper only the laboratory reaction rates are discussed since in the measurement the target nucleus is always in the ground state. In the case of the $^{92}$Mo(p,$\gamma$)$^{93}$Tc reaction, however, the difference between the laboratory and stellar reaction rates is always very small in the relevant temperature range \cite{NONSMOKER}.}  of a given reaction in a stellar plasma of temperature $T$ is given by the following formula \cite{ili07}:
\begin{equation}
\label{eq:rate}
	N_A<\sigma v> = N_A\left(\frac{8}{\pi m} \right)^{1/2}\frac{1}{(kT)^{3/2}}\int^{\infty}_{0}{E\sigma(E)e^{-E/kT}}dE \equiv \int^{\infty}_{0}\sigma(E)W(E)dE
\end{equation}
where $N_A$ is the Avogadro number, $m$ is the reduced mass of the reactants, $\sigma(E)$ is the cross section, $k$ is the Boltzmann constant and $E$ is the energy which runs over the Maxwell-Boltzmann energy distribution of the reacting nuclei. For the calculation of the reaction rate the cross section as a function of energy is needed. If, however, the thick target yield as a function of energy is known in the astrophysically relevant energy range, the intermediate step of cross section determination can be avoided \cite{rou76}. By differentiating Eq.\,(\ref{eq:yield}) and substituting into Eq.\,(\ref{eq:rate}) one gets:

\begin{equation}
\label{eq:ratefromyield}
	N_A<\sigma v> =  \int^{\infty}_{0}\frac{dY(E)}{dE}\varepsilon(E)W(E)dE = \int^{\infty}_{0}Y(E)\frac{d\left(\varepsilon(E)W(E)\right)}{dE}dE
\end{equation}
where the second formula was obtained by integration by parts. The $\varepsilon(E)$ stopping power is known as well as the $W(E)$ function (defined by Eq.\,(\ref{eq:rate})) which contains the energy dependence of the Maxwell-Boltzmann distribution. The differentiation can thus be done. 

The $Y(E)$ thick target yield is measured in a finite energy range. The integration, however, must be carried out from zero to infinite energy. Therefore, the thick target yield function must be extrapolated both to low and high energies. In the present work the extrapolation was done with exponential functions matching the value and slope of the measured yield functions at the two extreme energies, i.e. at 1700 and 3100 keV. 

\begin{landscape}
\begin{table}
\centering
\caption{\label{tab:rate} Thermonuclear reaction rate of the $^{92}$Mo(p,$\gamma$)$^{93}$Tc reaction. See text for details.}
\begin{tabular}{cr@{\,}c@{\,}lccc|cr@{\,}c@{\,}lc}
\hline
T & \multicolumn{3}{c}{Present work} & \multicolumn{3}{c}{Rate contribution} & Sauter and K\"appeler \cite{sau97} & \multicolumn{3}{c}{Roughton \textit{et al.} \cite{rou79}} & NON-SMOKER \cite{NONSMOKER}\\
\cline{5-7}
\cline{8-12}
\,[GK] & \multicolumn{3}{c}{[cm$^3$s$^{-1}$mole$^{-1}$]} & Low & Exp. & High & \multicolumn{5}{c}{[cm$^3$s$^{-1}$mole$^{-1}$]}\\

\hline
1.0	&	\multicolumn{3}{c}{(1.6	$\pm$	1.2)$\times$10$^{-4}$}	&	72\%	&	28\%	&	0\%	&	N/A & \multicolumn{3}{c}{N/A} & 1.8$\times$10$^{-4}$	\\
1.5	&	0.040	&	$\pm$	&	0.011	&	21\%	&	79\%	&	0\%	&	0.055 & \multicolumn{3}{c}{N/A} & 	0.051	\\
2.0	&	1.08	&	$\pm$	&	0.19	&	7\%	&	90\%	&	3\%	&	1.49 & 1.45& $\pm$	&	0.49 & 	1.44	\\
2.5	&	9.11	&	$\pm$	&	1.77	&	3\%	&	88\%	&	9\%	&	13.2 & 10.5& $\pm$	&	1.6 & 	13.8	\\
3.0	&	40.5	&	$\pm$	&	10.6	&	1\%	&	81\%	&	18\%	&	65.7 & 46.5& $\pm$	&	7.0 & 	72.8	\\
3.5	&	122	&	$\pm$	&	42	&	1\%	&	72\%	&	27\%	&	226 & 151& $\pm$	&	23 & 	265	\\
4.0	&	283	&	$\pm$	&	120	&	0\%	&	63\%	&	37\%	&	612 & 392& $\pm$	&	59 & 	752	\\
4.5	&	549	&	$\pm$	&	271	&	0\%	&	55\%	&	45\%	&	N/A & 865& $\pm$	&	130 & 	1790	\\
\hline
\end{tabular}
\end{table}
\end{landscape}

The obtained reaction rate values are listed in Table~\ref{tab:rate}. The columns labeled ``Rate contribution'' indicate how much the different energy regions contribute to the rate. The column labeled ``Exp.'' shows the contribution of the 1700\,--\,3100\,keV experimentally studied energy range to the rate. Similarly, columns ``Low'' and ``High'', respectively, show the contribution of low and high energy yield extrapolation to the rate. 

The table clearly shows that in the temperature range between about 1.5 and 3.5 GK the experimentally determined yields give the dominant contribution to the reaction rate. This energy range is relevant for the $\gamma$-process nucleosynthesis, therefore almost purely experimental reaction rate can be provided for this reaction.

The uncertainty of the reaction rate is calculated by the following procedure: in the measured 1700\,--\,3100\,keV energy range the experimental yield uncertainty of typically 7\,\% is used. Outside this region, both to high and low energies, a 100\,\% uncertainty is assigned to the extrapolated yield. The uncertainty values listed in Table~\ref{tab:rate} reflects this procedure: where the rate is determined by the experimental energy range the uncertainty is relatively low, while it is increasing towards lower and higher temperatures. This is the reason why the rate values are shown only in the 1.0\,--\,4.5\,GK temperature interval. 

The last columns of the table show the rates obtained by Sauter and K\"appeler \cite{sau97}, Roughton \textit{et al.} \cite{rou79} and calculated with the NON-SMOKER code \cite{NONSMOKER}. The results of Roughton \textit{et al.} are in good agreement with our data while the rates of Sauter and K\"appeler are higher by about a factor of 1.4 to 2.2. 

The comparison with the statistical calculations carried out with the NON-SMOKER code shows that the model overestimates the reaction rate by about a factor of 2. This is in agreement with the S-factor comparison which is shown in Fig.\,\ref{fig:92totMo_res}. The model overestimates the experimental cross section and the deviation increases with increasing energy. This explains the increasing difference between the measured and calculated rates towards higher temperatures. 

\section{Summary}

The $^{92}$Mo(p,$\gamma$)$^{93}$Tc and $^{98}$Mo(p,$\gamma$)$^{99m}$Tc reaction were studied by measuring thick target yields in the proton energy interval between 1700 and 3100 keV. By differentiating the thick target yield curve, cross sections of the studied reactions were calculated. In the case of the $^{98}$Mo(p,$\gamma$)$^{99m}$Tc reaction the obtained cross sections are in good agreement with previous data and model calculations.

The previously observed strong fluctuations in the $^{92}$Mo(p,$\gamma$)$^{93}$Tc reaction cross section are confirmed in the present work. For this reaction partial cross sections leading to the ground and isomeric state in $^{93}$Tc were measured separately. The isomeric state cross section is in good agreement with the results of Sauter and K\"appeler \cite{sau97}, while the obtained ground state cross sections are lower than the Sauter and K\"appeler \cite{sau97} data and are in better agreement with the preliminary results of the Cologne group \cite{has10,sau11,sau11b}. 

We have argued that in the case of a reaction with strongly fluctuating cross section it is more useful for astrophysical applications to measure thick target yield than differential cross section with thin targets. Without the intermediate step of cross section determination we have derived astrophysical reaction rate for the $^{92}$Mo(p,$\gamma$)$^{93}$Tc reaction. In the temperature range relevant for the $\gamma$-process the reaction rate is predominantly determined by our experimental data. Our new recommended reaction rates are about a factor of 2 lower than the rates provided by the NON-SMOKER code which is widely used in astrophysical network calculations.

\section*{Acknowledgments}

This work was supported by ERC Grant 203175, and OTKA NN83261 (EuroGENESIS), K108459, K101328, PD104664. G.G.K. is a Bolyai fellow.


\begin{thebibliography}{00}


\bibitem{kap11} F. K\"appeler, R. Gallino, S. Bisterzo, W. Aoki, Rev. Mod. Phys. 83 (2011) 157.
\bibitem{arn07} M. Arnould, S. Goriely, K. Takahashi, Phys. Rep. 450 (2007) 97.
\bibitem{rau13} T. Rauscher, N. Dauphas, I. Dillmann, C. Fr\"ohlich, Zs. F\"ul\"op, Gy. Gy\"urky, Rep. Prog. Phys. 76 (2013) 066201.
\bibitem{KADONIS} T. Sz\"ucs, I. Dillmann, R. Plag, Zs. F\"ul\"op, J. Phys.: Conf. Series 337 (2012) 012033.
\bibitem{rau11} T. Rauscher, J. Mod. Phys. E 20 (2011) 1071.
\bibitem{sau97} T. Sauter, F. K\"appeler, Phys. Rev. C 55 (1997) 3127.
\bibitem{arn03} M. Arnould, S. Goriely, Phys. Rep. 384 (2003) 1.
\bibitem{has10} J. Hasper, M. B\"ussing, M. Elvers, J. Endres, A. Zilges, J. Phys. Conf. Ser. 202 (2010) 012005.
\bibitem{sau11} A. Sauerwein, M. Elvers, J. Endres, J. Hasper, A. Hennig, L. Netterdon, A. Zilges, AIP Conf. Proc. 1377 (2011) 423.
\bibitem{sau11b} A. Sauerwein, M. Elvers, J. Endres, J. Hasper, A. Hennig, L. Netterdon, A. Zilges, Proceedings of Science, PoS(NIC XI) (2011) 244.
\bibitem{SRIM-Mo} http://www.srim.org/SRIM/SRIMPICS/STOP01/STOP0142.gif
\bibitem{NDS92} C.M. Baglin, Nucl. Data Sheets 112 (2011) 1163.
\bibitem{NDS98} E. Browne, J.K. Tuli, Nucl. Data Sheets 112 (2011) 275.
\bibitem{rau10} T. Rauscher, Phys. Rev. C 81 (2010) 045807.
\bibitem{hal12} Z. Hal\'asz et al., Phys. Rev. C 85, (2012) 025804.
\bibitem{rou79} N.A. Roughton, M.R. Fritts, R.J. Peterson, C.S. Zaidins, C.J. Hansen, At. Data Nucl. Data Tables 23 (1979) 177.
\bibitem{TALYS} A.J. Koning \textit{et al.}, AIP Conf. Proc. 769 (2005) 1154.
\bibitem{kon03} A.J. Koning and J.P. Delaroche, Nucl. Phys. A713 (2003) 231.
\bibitem{bri57} D.M. Brink, Nucl. Phys. 4 (1957) 215.; P. Axel, Phys. Rev. 126 (1962) 671.
\bibitem{gil65} A. Gilbert and A.G.W. Cameron, Can. J. Phys. 43 (1965) 1446.
\bibitem{ili07} C. Iliadis, Nuclear Physics of Stars, WILEY-VCH Verlag GmbH \& Co. KGaA, 2007.
\bibitem{NONSMOKER} T. Rauscher, F-K. Thielemann, At. Data Nucl. Data Tables 79 2001 47., www.nucastro.org
\bibitem{JLM} J.P. Jeukenne, A. Lejeune, C. Mahaux, Phys. Rev. C 16 (1977) 80.
\bibitem{rau97} T. Rauscher, F-K. Thielemann, K-L. Kratz, Phys. Rev. C 56 (1997) 1613.
\bibitem{NDS92old} H. Sievers, Nucl. Data Sheets 54 (1988) 99.
\bibitem{rou76} N.A. Roughton, M.J. Fritts, R.J. Peterson, C.S. Zaidins, C.J. Hansen, Astrophysical Journal 205, (1976) 302. 

\end{thebibliography}
\end{document}